\documentclass[aps,pra,twocolumn,10pt,superscriptaddress,showpacs,amsmath,amssymb,nofootinbib]{revtex4-1}

\usepackage{graphicx}
\usepackage{hyperref}
\usepackage{threeparttable}

\newcommand{\demtab}{
\begin{table}[b]
\begin{ruledtabular}
\caption{Comparison of student demographics collected via survey, advising sheets, and departmental data for all majors.}
\label{table1}
\begin{tabular}{lccc}
& Survey & Advising & All majors \\ 
& $N=76$ & $N=394$ & $N=493$ \\ \hline
Physics majors & 55\% & 66\% & 71\% \\
Engingeering physics majors & 43\% & 34\% & 29\% \\
URE (Fall 2014 only)& 50\% & 23\% & -- \\
URE (Fall 2014 or before) & 59\% & -- & -- \\
Women & 16\% & -- & 12\% \\
Underrepresented minorities & 5\% & -- & 8\% \\
\end{tabular}
\end{ruledtabular}
\end{table}
}

\newcommand{\datafig}{
\begin{figure}[t]
\centering
\includegraphics{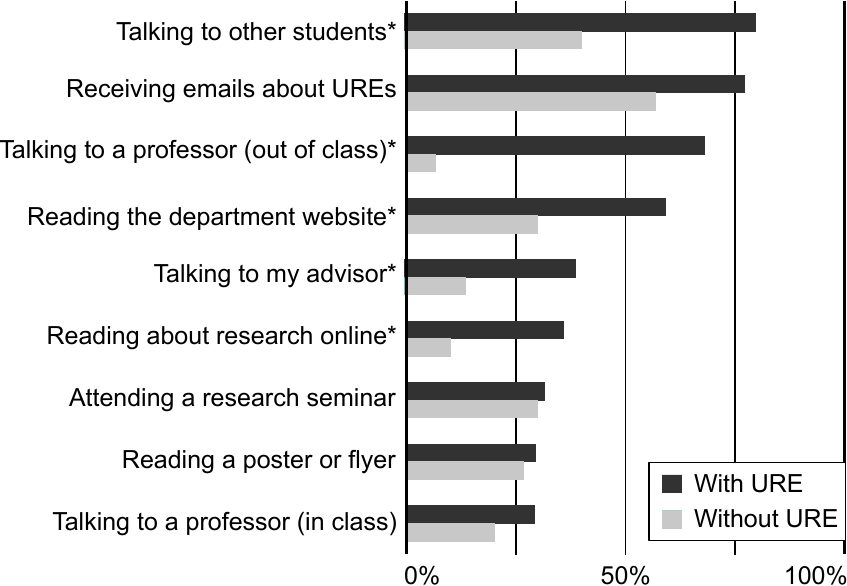}
\caption{Sources of information about UREs, broken down by students with research experience and those without. Asterisks indicate items on which differences between the two groups were statistically significant. Items selected by fewer than 20\% of both groups are not shown.}
\label{sources}
\end{figure}
}

\newcommand{\datafigmentor}{
\begin{figure}[b]
\centering
\includegraphics{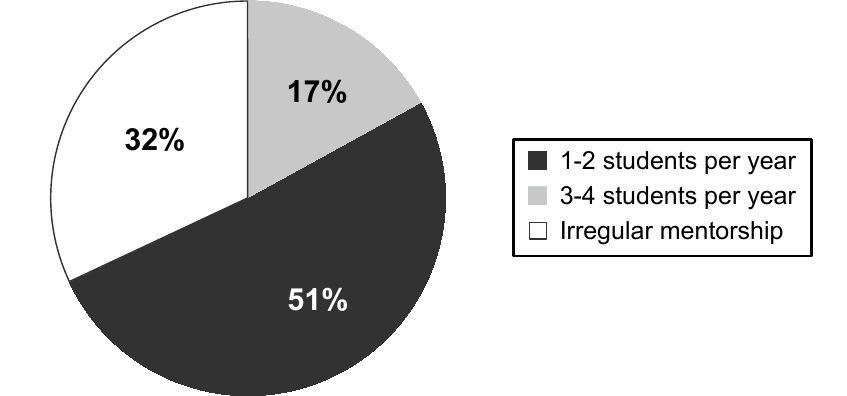}
\caption{Typical number of undergraduate students mentored annually, per faculty member.  ``Irregular mentorship" means faculty members may not mentor undergraduate researchers every year.}
\label{facultycap}
\end{figure}
}

\newcommand{\datafigtasks}{
\begin{figure}[t]
\centering
\includegraphics{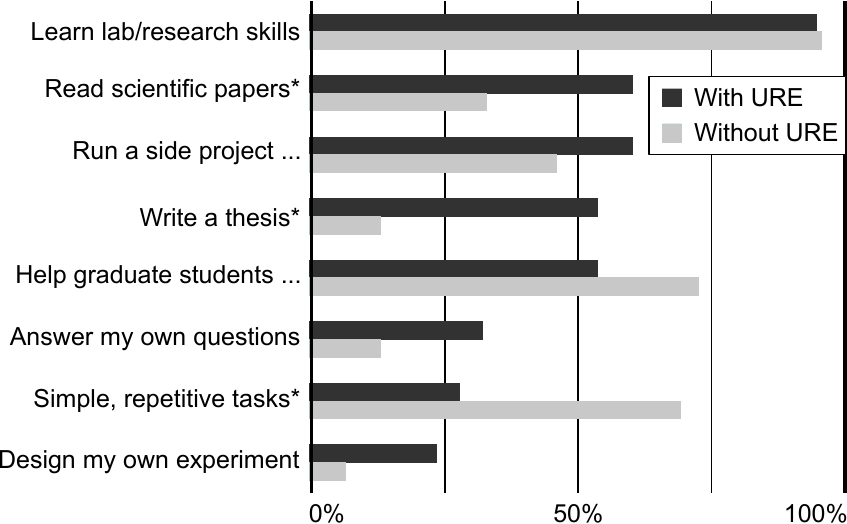}
\caption{Perceptions of research practices among students with research experience ($N=46$) and those without ($N=30$). Asterisks indicate items on which differences between the two groups were statistically significant. Items selected by fewer than 20\% of both groups are not shown.}
\label{tasks}
\end{figure}
}

\begin{document}

\title{Access to undergraduate research experiences at a large research university}

\author{S. 5 Hanshaw}
\affiliation{Department of Physics, University of Colorado, Boulder, CO 80309, USA}

\author{Dimitri R. Dounas-Frazer}
\affiliation{Department of Physics, University of Colorado, Boulder, CO 80309, USA}

\author{H. J. Lewandowski}
\affiliation{Department of Physics, University of Colorado, Boulder, CO 80309, USA}
\affiliation{JILA, National Institute of Standards and Technology and University of Colorado, Boulder, CO, 80309, USA}

\pacs{01.40.Fk}


\begin{abstract}
The American Physical Society recently released a statement calling on all university physics departments to provide or facilitate access to research experiences for all undergraduate students. In response, we investigated the current status of access to undergraduate research at University of Colorado Boulder (CU), a large research institution where the number of undergraduate physics majors outnumber faculty by roughly ten to one. We created and administered two surveys within CU's Physics Department: one probed undergraduate students' familiarity with, and participation in, research; the other probed faculty members' experiences as research mentors to undergraduates. We describe the development of these instruments, our results, and our corresponding evidence-based recommendations for improving local access to undergraduate research experiences. Reflecting on our work, we make several connections to an institutional change framework and note how other universities and colleges might adapt our process.
\end{abstract}

\maketitle

\section{Introduction}

Undergraduate research experiences (UREs) have many benefits for students and faculty members. Student benefits include improved technical skills, a deeper understanding of the nature of science, and a stronger sense of identity as a scientist~\cite{russell_benefits_2007,taraban_academic_2012,sandra_laursen_undergraduate_2010,bunnet_education_1984}. Benefits to research mentors include improved intellectual maturity~\cite{khoukhi_structured_2013}. The many positive impacts of UREs inform a long history of national pressure to increase student participation in research~\cite{shirley_strum_kenny_reinventing_1998,business-higher_education_forum_u.s._2013,presidents_council_of_advisors_on_scienc_and_technology_engage_2012}. Most recently, the American Physical Society (APS) released a statement encouraging ``the nation's four-year colleges and universities and their physics and astronomy departments to provide or facilitate access to research experiences for all undergraduate physics and astronomy majors"~\cite{american_physics_society_council_undergraduate_2014}.

There are many challenges associated with engaging all students in UREs, such as decreases in mentors' research productivity \cite{adedokun_exploring_2010}, infrastructural constraints imposed by lack of lab space, high student-to-mentor ratios, and other issues~\cite{desai_integrating_2008}. Importantly, the APS calls for providing undergraduate students with \emph{access to research}, which is a broader goal than providing \emph{research opportunities}. In the present work, we explore what it means to facilitate access to UREs in the University of Colorado Boulder (CU) Physics Department. We define ``access" as consisting of two pieces: (1) \emph{awareness} of the existence of UREs, what they entail, why UREs are beneficial educational experiences, and how to apply for them; and (2) \emph{selection} from among a variety of available on- and off-campus research opportunities. 

We describe the current state of access to UREs in the Physics Department at CU and present recommendations to improve access beyond the status quo. We worked with both students and faculty members in order to build sustainable efforts to improve access to UREs. Because this work represents a first step in a department-wide change effort, we reflect on our process through multiple perspectives on institutional change~\cite{corbo_sustainable_2014}. Our process and results may be applicable to other large research institutions working towards similar goals.

\section{Institutional Context}
The CU Physics Department is among the largest in the country in terms of number of bachelor's degrees awarded annually~\cite{mulvey_physics_2012}. Despite being an active research university, only about 20\% of physics and engineering physics students (collectively, ``majors") were actively engaged in UREs during Fall 2014. This rate of undergraduate engagement in research is partially due to an infrastructural constraint: in a department with about 500 majors, the student-to-faculty ratio is 10:1, making it difficult to provide on-campus UREs to all students.

As we discuss in more detail below, one strategy for improving access to UREs involves connecting students to off-campus research experiences. In particular, CU is near to a number of research facilities, such as: National Center for Atmospheric Research (NCAR), National Institute of Standards and Technology (NIST), and National Oceanic and Atmospheric Administration (NOAA) (collectively, ``national labs"). Though these national labs are not part of CU, we consider their close proximity to the university an important local resource.

\section{Methods}

To probe the current state of access to UREs at CU, we collected interview and survey data from faculty members and majors in the CU Physics Department during Fall 2014. Student participants were paid volunteers; faculty members were not paid. In total, we conducted interviews with 7 students and 2 faculty members. Interview data were used only to inform survey design and are not elaborated herein. Survey participants included 47 of 50 total faculty members (94\% response rate) and 76 of 493 majors (15\% response rate). Given the relatively low student response rate, we compared student participant demographics to two external sources of demographic information: departmental data on majors' gender, race, and ethnicity; and majors' participation in UREs during the Fall 2014 semester, as reported on advising forms.

We created and administered two surveys: one for faculty members and another for students. The faculty survey consisted of 25 questions focusing on mentoring capacity, expectations of students, and desired help from the department. The student survey consisted of 35 questions focusing on awareness of UREs and research-related programs, perceptions of UREs, and experiences in research. Both surveys included forced-choice, multiple-response, and open-ended items. Herein, we limit our discussion to responses to forced-choice and multiple-response questions. When discussing student results, our analysis focused on comparing responses between students with research experience and those without. When appropriate, we used the Chi Squared test~\cite{chi} to determine statistical significance in responses. 

\section{Results}

\demtab

As mentioned above, only 15\% of majors completed the student survey on UREs. Demographic comparisons of student participants to the population of majors are presented in Table~\ref{table1}. In our sample, two groups were overrepresented compared to the student population as a whole: (1) engineering physics majors and (2) students actively engaged in research during Fall 2014. Differences in representation of women and underrepresented minority students were not statistically significant.

One potential explanation for overrepresentation of engineering physics majors in our sample is that one of us (H.J.L.) is the faculty advisor for this group of students; for some students, this relationship may have increased motivation to complete the survey. On a similar note, students actively participating in research may have been more motivated to complete a survey on UREs than their non-participating counterparts. For example, students with no research experience may have felt that a survey about UREs didn't apply to them.

Due to the overrepresentation of students engaged in UREs at the time the survey was administered, student survey participants are likely more aware of the existence and nature of UREs than the typical major. Therefore, our data likely bias our understanding of barriers to access to UREs. However, the barriers we were able to identify may be exacerbated among students who have not previously engaged in research. In order to reduce these barriers, we identified ways to improve access by increasing awareness and selection of UREs. 


Compared to students without research experience, those with research experience were more familiar with research-related programs. By ``research-related programs," we mean departmental research seminars targeted at an undergraduate audience, campus-wide apprenticeship programs that provide stipends for on-campus UREs, and national programs like the NSF REU program. Similarly, students with research experience were more familiar with the different research groups in the Physics Department as well as local national labs. Though it is unclear whether or how these differences in familiarity are causally connected to participation in UREs, we nevertheless identify the following area for improvement: \emph{raise awareness of research-related programs, research groups, and local national labs among majors}.

\datafigtasks

Further, students with research experience were more likely to identify tasks that scientists do, such as ``reading scientific papers" and ``writing a thesis", as parts of the research experience, whereas students without research experience focused on more assistant-type aspects of a URE, like ``simple, repetive tasks" and ``helping graduates" (Fig.~\ref{tasks}). This difference informs our second area for improvement: \emph{raise awareness of the practices of undergraduate research}.

Regardless of research experience, students identified a variety of means of learning about UREs, including: talking to professors or other students, reading posters or flyers, and attending research seminars (Fig.~\ref{sources}). In general, students without research experience were more likely to report learning about UREs from fewer sources than researchers. Given the prevalence of informal, decentralized sources of information about UREs in the Physics Department, we identified a third area for improvement: \emph{raise awareness about pathways to UREs through official and centralized communication channels}.

\datafig

Additionally, 42\% of students with research experience indicated that they would have liked to start their URE earlier in their undergraduate career, particularly those who had started after their second year.  This finding aligns well with national calls to engage undergraduate students in research as early as their first year of college~\cite{shirley_strum_kenny_reinventing_1998}. However, among students with research experience, only about a third (38\%) began conducting research within their first year at CU. Therefore, we concluded that our recommendations must serve to \emph{raise awareness and promote positive student perceptions about UREs as early in their undergraduate career as possible}. 

Finally, the faculty survey revealed two constraints on increasing selection of UREs. First, about half of faculty members mentor 1--2 students per year (Fig.~\ref{facultycap}). A majority of both students and faculty members indicated a preference for UREs lasting at least a year, consistent with recommendations in the literature~\cite{russell_benefits_2007}. Together, the low number of students mentored and the desire for experiences that span multiple semesters limit the number of UREs that can be offered in the department. Second, half of the faculty members indicate that they would have made no change to their most recent mentoring experience; faculty members commented that the department should provide funding, student skill-building, and networking assistance rather than interfering with faculty members' autonomy in mentoring. Thus, we decided that any recommendations we made should \emph{avoid changing the nature of the mentor-student relationship, including the number of students mentored by each faculty member}. 

\section{Recommendations}

\datafigmentor

In response to the areas of improvement which emerged from the survey data, we crafted five recommendations for improving students access to UREs in the CU Physics Department: (1) create a ``Frequently-Asked-Questions (FAQ) About UREs" page on the department's website, (2) host a symposium on access to UREs each fall, (3) host a poster session for UREs each spring, (4) develop a sophomore-level Research Methods course, and (5) explore partnerships with local national labs to create opportunities for off-campus UREs.

The FAQ page and symposium may improve access to UREs by using official, centralized channels to raise awareness about research-related programs, diverse pathways to obtaining UREs, on- and off-campus experiences, and so on. The poster session and methods course, on the other hand, may raise awareness about the common research practices of undergrads involved in UREs. Having the symposium in the fall semester and targeting an audience of first-year students will help inform students about UREs and encourage them to seek out opportunities early in their undergraduate careers. Together, these four recommendations provide complementary support that address all areas for improvement. Exploring partnerships with local national labs may help increase the selection of off-campus UREs available to our students while honoring the constraint that faculty members do not want to change their mentoring practices.

A first draft of the recommendations was circulated among faculty members, resulting in minor feedback. Multiple faculty members committed to publicly supporting the recommendations. A revised draft was presented at a departmental meeting in Spring 2015. The recommendations received support from the department, and a Committee on Undergraduate Research was created to oversee their implementation.

\section{Change Perspectives}

Working towards the APS call to provide all majors with access to UREs requires department-wide changes. Accordingly, we reflected on our process through the framework for institutional change outlined by Corbo et al.~\cite{corbo_sustainable_2014}. In particular, we focused on two theoretical perspectives that typically underlie change efforts: a \emph{social cognition perspective}, which involves identifying and potentially changing the underlying beliefs of a group; and a \emph{political perspective}, which includes building coalitions to support collective actions. 

From a social cognition perspective, our surveys helped identify the relevant underlying beliefs of both faculty members and students. Our recommendations were designed to change some of these beliefs while honoring others. For example, the poster session and methods course may challenge students' beliefs about the nature of UREs, allowing them to make better-informed decisions about whether or not to pursue such experiences. However, we chose to align with faculty members' belief that their mentor-student relationships should not be changed in a significant way. One way to honor these underlying beliefs was to recommend looking off-campus for additional opportunities.

Our strategy for building support for our recommendations aligns well with a political perspective. By engaging individual faculty members in discussion about our recommendations before making an official proposal to the department, we were able to establish a supportive coalition for our ideas. Our proposal resulted in the creation of an official departmental committee that has been tasked with taking collective action towards implementing our recommendations.

Reflecting on our process through the lens of institutional change is also useful for informing future efforts. For example, it may be useful to unpack underlying beliefs about the nature of the mentor-student relationship to develop better ideas about whether and how those relationships might be supported and improved.

\section{Summary and Future Directions}

As an initial step towards meeting the APS call to provide all majors with access to UREs, we conducted student and faculty surveys to establish a baseline understanding of access to UREs in the CU Physics Department. Survey results informed a set of five recommendations for improving access to UREs at CU. Finally, we used a framework for institutional change to reflect on both our process and results, focusing specifically on social cognition and political perspectives on change. Future work will focus on implementation and evaluation of our recommendations at CU. Other institutions just beginning to answer the APS call may find our process useful in addressing access to UREs for their own faculty members and undergraduates. 

The institutional change framework helps clarify whether and how our process and results may be translated to institutional contexts other than the CU Physics Department. Identifying faculty members' and students' underlying beliefs about access to UREs, developing recommendations that explicitly address those beliefs (either by aligning with or changing them), and building coalitions of support for the recommendations are practices that can be applied at any institution. Additionally, the instruments developed at CU and the corresponding recommendations for change may be useful to other research institutions whose faculty members and students have similar underlying beliefs about UREs.

\acknowledgments The authors acknowledge Joel Corbo and the rest of the CU PER Group for useful discussions. This work was supported by the Research Corporation for Science Advancement and NSF grant no. DUE-1323101.


%

\end{document}